\documentclass[unnumsec,webpdf,contemporary,large]{oup-authoring-template}%
\graphicspath{{Fig/}}
\usepackage{booktabs}
\usepackage{tabularx}
\usepackage{array}
\usepackage{booktabs}
\usepackage{graphicx}
\usepackage{ragged2e}

\theoremstyle{thmstyleone}%
\theoremstyle{thmstyletwo}%
\theoremstyle{thmstylethree}%
\begin{document}

\journaltitle{Journal Title Here}
\DOI{DOI added during production}
\copyrightyear{YEAR}
\pubyear{YEAR}
\vol{XX}
\issue{x}
\access{Published: Date added during production}
\appnotes{Paper}

\firstpage{1}

%\subtitle{Subject Section}

\title[Short Article Title]{Time-Varying Environmental and Polygenic Predictors of Substance Use Initiation in Youth: A Survival and Causal Modeling Study in the ABCD Cohort}

\author[1,$\ast$]{Mengman Wei\ORCID{0009-0008-0458-7140}}
\author[1, ]{Qian Peng}

\address[1]{\orgdiv{Department of Neuroscience}, \orgname{The Scripps Research Institute}, \orgaddress{\street{10550 N Torrey Pines Rd, La Jolla}, \postcode{92037}, \state{CA}, \country{U.S.}}}

\corresp[$\ast$]{Corresponding author. \href{email:email-id.com}{mwei@scripps.edu}}

\received{Date}{0}{Year}
\revised{Date}{0}{Year}
\accepted{Date}{0}{Year}

%\editor{Associate Editor: Name}

%\abstract{
%\textbf{Motivation:} .\\
%\textbf{Results:} .\\
%\textbf{Availability:} .\\
%\textbf{Contact:} \href{name@email.com}{name@email.com}\\
%\textbf{Supplementary information:} Supplementary data are available at \textit{Journal Name}
%online.}

\abstract{Early initiation of alcohol, nicotine, cannabis, and other substances is associated with elevated risk of later substance use disorders and related psychiatric outcomes. However, the relative contributions of dynamic environmental exposures, behavioral risk factors, and genetic liability to the timing of substance use initiation during early adolescence remain unclear. We integrated longitudinal environmental measures with polygenic risk scores in the Adolescent Brain Cognitive Development Study to identify robust predictors of substance use initiation and to evaluate selected time-varying predictors using marginal structural models.\\
We analyzed unrelated participants of European genetic ancestry from the ABCD Study with available longitudinal substance use data, environmental measures, covariates, and polygenic risk scores. Four time-to-event outcomes were defined: alcohol initiation, nicotine initiation, cannabis initiation, and any substance use initiation through approximately four years of follow-up. Time-varying start--stop interval datasets were constructed using interview age in months. We first fitted covariate-adjusted univariate time-varying Cox models to screen candidate predictors. We then fitted LASSO-selected multivariable time-varying Cox models adjusted for age, sex, ancestry principal components, and study site. Polygenic risk scores for alcohol, cannabis, nicotine, and general substance use disorder were included as candidate genetic predictors. Finally, marginal structural models with inverse probability of treatment weighting were applied to selected modifiable time-varying predictors to assess the robustness of associations under a causal modeling framework.\\
The final analytic cohort included 2,366 EUR unrelated participants, with initiation events observed for alcohol, nicotine, cannabis, and any substance use. Univariate time-varying Cox models identified broad associations across behavioral, family, socioeconomic, psychosocial, lifestyle, and environmental domains. In multivariable Cox models, the strongest and most consistent predictors were environmental and behavioral factors, including youth rule-breaking behavior, impulsivity-related traits, parental alcohol-related problems, financial adversity, family and cultural context, media or phone use, life-event burden, peer or romantic experiences, sleep-related factors, and parental monitoring. In contrast, polygenic risk scores showed weaker and less consistent evidence of association. Although some PRS associations appeared in screening-level models, they did not emerge as robust predictors after stringent correction, multivariable selection, or downstream causal modeling. In marginal structural models, youth rule-breaking behavior was the most consistent predictor, showing robust associations with alcohol, nicotine, cannabis, and any substance initiation. Additional robust or outcome-specific MSM signals included sensation seeking, lack of planning, parental alcohol-related problems, parent-rated child rule-breaking symptoms, youth phone use, negative life-event affect, romantic relationship experience, caffeine exposure, weekday sleep duration, family obligation values, and parental monitoring.\\
Early substance use initiation in adolescence was more strongly associated with dynamic environmental, behavioral, family, and psychosocial factors than with measured common-variant genetic liability in this analytic cohort. The most robust signals pointed to potentially modifiable developmental pathways, particularly behavioral dysregulation, impulsivity, parental substance-related problems, family and socioeconomic context, peer or romantic experiences, media use, sleep, caffeine exposure, and parental monitoring. These findings support prevention strategies focused on proximal environmental and behavioral risk contexts, while suggesting that PRS may be better interpreted as background genetic liability rather than as central robust predictors of early initiation in this study.
}

\keywords{adolescence; ABCD Study; substance use initiation; time-varying Cox model; marginal structural model; inverse probability weighting; environmental risk factors; polygenic risk score}

\maketitle

%\begin{epigraph}
%Epigraph text. Ximporem qui reperov idempedit modio. Bisto imagnatem quae aceptis
%nobitae quid eum rae adignis quias-sit vellacc uptatur sunt quis rentis eaquasit alia deliquam
%rec-to consed unt. Empor sum ratur ressimusdae. Nam fugiae.
%\source{Epigraph source}
%\end{epigraph}

\section{Introduction}

Early initiation of alcohol, nicotine, cannabis, and other substances is consistently associated with increased risk of substance use disorders, polysubstance use, and broader externalizing and internalizing psychopathology in later adolescence and adulthood~\cite{1, 2, 3, 4}. Understanding which environmental and genetic factors shape the timing of initiation is a critical step toward designing effective preventive interventions.

The Adolescent Brain Cognitive Development (ABCD) Study®~\cite{5, 6, 7} provides a unique opportunity to study the development trajectory of these initiations in youth. ABCD follows a large, diverse cohort of U.S. youth with repeated assessments of substance use, demographic and socioeconomic context, family and peer functioning, neighborhood characteristics, mental and physical health, cognition, and neuroimaging, alongside genome-wide genotyping. This allows us to study substance use initiation as a time-to-event outcome under rich longitudinal covariate information.

\begin{figure*}[htbp]
\centering
\includegraphics[width=\textwidth]{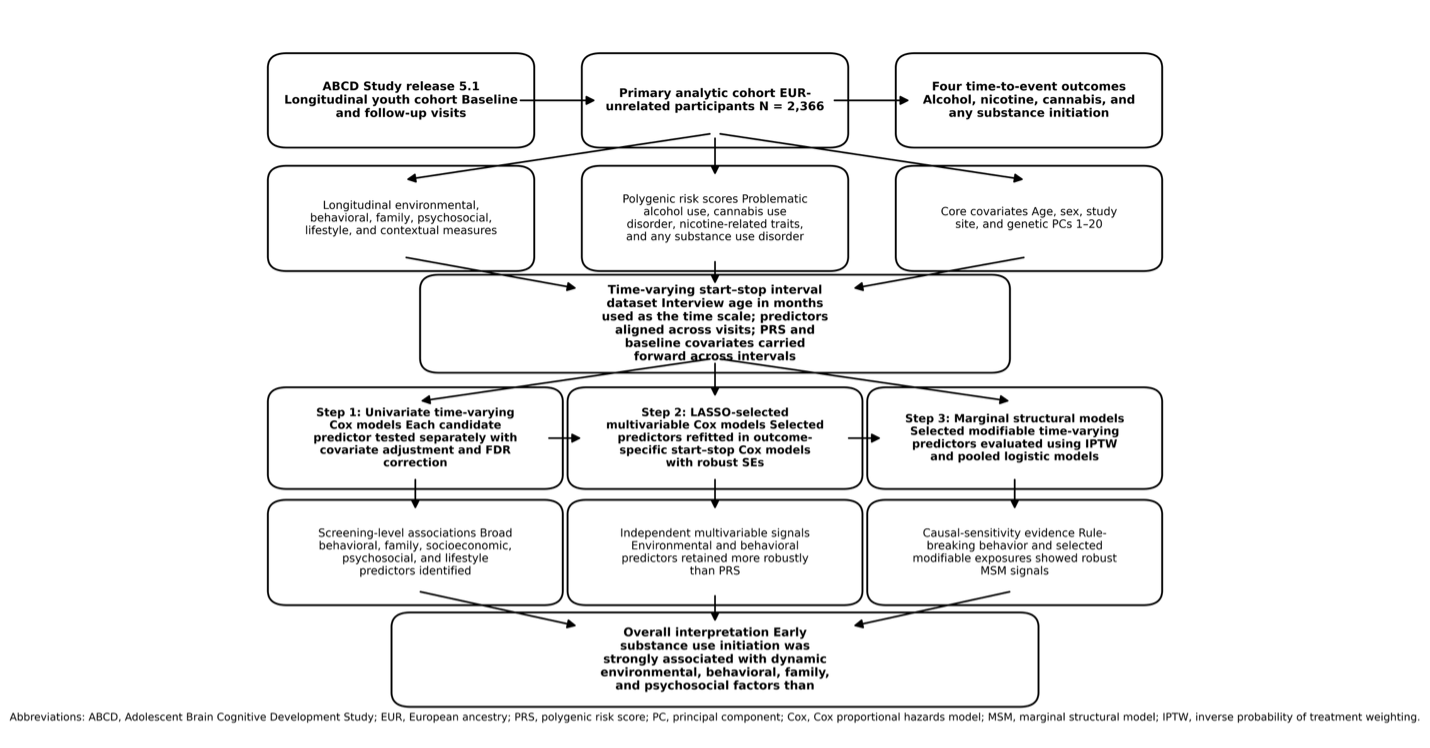}
\caption{Study design and analytic workflow. The workflow summarizes construction of the EUR-unrelated ABCD analytic cohort, definition of four substance use initiation outcomes, integration of longitudinal environmental and behavioral measures with polygenic risk scores and core covariates, construction of time-varying start–stop survival datasets, and the three-stage analytic strategy. Candidate predictors were first screened using covariate-adjusted univariate time-varying Cox models, followed by LASSO-selected multivariable time-varying Cox models and marginal structural models with inverse probability of treatment weighting for selected modifiable predictors. The overall analysis evaluated whether early substance use initiation was more strongly associated with dynamic environmental and behavioral factors or with measured common-variant genetic liability.}
\label{Figure1}
\end{figure*}

In this work (Figure~\ref{Figure1}), we focus on two main aims:
Aim 1 (Association): To identify time-varying environmental and polygenic predictors of substance use initiation (alcohol, nicotine, cannabis, and any substance) using time-varying Cox proportional hazards models~\cite{8}.

Aim 2 (Causal exploration): To perform focused causal analyses for a subset of modifiable environmental predictors that show robust associations in Aim 1, using marginal structural models~\cite{9, 10} with inverse probability of treatment weighting.

\section{Methods}
\subsection{Study Cohort}

We used data from the Adolescent Brain Cognitive Development (ABCD) Study\textsuperscript{\textregistered} release 5.1~\cite{5}. Participants were included if they had genome-wide genotyping data passing quality control, at least one assessment of substance use, and complete data for core covariates (sex, age, study site, and ancestry principal components). The analytic sample comprised 11{,}868 children from the ABCD baseline cohort (mean baseline age: $9.91 \pm 0.62$ years). Longitudinal follow-up data were available for approximately four years.

Substance use initiation was defined separately for alcohol, nicotine, cannabis, and any substance~\cite{11}. Time-to-event variables were constructed as the number of months from baseline to first reported use. Participants who did not report initiation were censored at their last available follow-up assessment. By the end of follow-up, initiation events were observed for alcohol (36.5\%), nicotine (5.44\%), cannabis (3.42\%), and any substance (39.7\%). Detailed distributions of initiation timing, as well as stratified summaries by sex and race/ethnicity, are provided in the Supplementary Materials.

\subsection{Time-varying covariates and candidate predictors}

We constructed time-varying environmental covariates from the Adolescent Brain Cognitive Development (ABCD) Study\textsuperscript{\textregistered} curated core tables. For each participant, we merged all available long-format variables across visits from multiple domains, including demographics and socioeconomic status, culture and environment, gender identity and sexual health, linked external data, mental health, neurocognition, novel technologies, and physical health. For each subject visit, these variables were aligned with the ABCD longitudinal tracking file to obtain interview age, and follow-up time was computed as months since baseline.

As adjustment covariates, we included sex, age at baseline, recruitment site, and the first 20 genetic principal components. These variables were retained in all models and treated as covariates rather than exposures of interest. In addition, we included four polygenic risk scores (PRS) as candidate predictors: problematic alcohol use (AUD)~\cite{12}, cannabis use disorder (CUD)~\cite{13}, nicotine-related traits (NUD)~\cite{14}, and any substance use disorder (anySUD)~\cite{15}.

We applied minimal filtering to candidate predictors to avoid discarding potentially informative but sparse or baseline-only variables. Specifically, we excluded (i) variables that were entirely missing, (ii) near-constant numeric variables, and (iii) highly collinear numeric variables (absolute pairwise correlation $> 0.9$). Baseline-only measures were retained and treated as time-varying covariates that remain constant across follow-up.

Polygenic risk scores were computed using external GWAS summary statistics and the PRS-CS~\cite{16} and PRS-CSx~\cite{17} pipelines. PRS were treated as time-invariant baseline covariates and carried forward across all time intervals for each participant. Detailed procedures for PRS construction are provided in the Supplementary Materials.

\subsection{Outcomes: time to substance use initiation}

For each ABCD participant, we defined four substance-use initiation outcomes (alcohol, nicotine, cannabis, and any substance) following logic similar to prior work~\cite{11}. Substance use was assessed using the Youth Substance Use Interview (annual in-person; \texttt{su\_y\_sui.csv}) and the Substance Use Phone Interview (mid-year; \texttt{su\_y\_mypi.csv}).

We harmonized these two data sources and ordered all observations by study visit (baseline, interim phone interviews between annual visits, and annual follow-ups). Follow-up was censored at the 4-year visit (FU4; \texttt{4\_year\_follow\_up\_y\_arm\_1}), such that only endorsements occurring on or before FU4 were considered initiation events. For each substance, initiation was defined as the first visit with any affirmative endorsement (coded as 1) among prespecified item sets capturing alcohol use (e.g., sip/full drink and TLFB alcohol use), nicotine use (e.g., cigarettes, e-cigarettes, cigars, chew, hookah, pipes, and other nicotine products), and cannabis use (e.g., puff/use, blunts, edibles, concentrates, tinctures, vaping oils). Responses were coerced to valid binary values (0/1), and invalid responses were treated as missing. To avoid misclassification of religious or ceremonial exposure as initiation, alcohol- and nicotine-related endorsements flagged as occurring exclusively in religious contexts were set to missing prior to aggregation. ``Any substance'' initiation was defined as initiation of alcohol, nicotine, or cannabis, and additionally included a broader set of ``other substance'' items (e.g., stimulants, opioids, hallucinogens, inhalants, and other drugs) to capture non-alcohol, non-nicotine, and non-cannabis initiation.

For each outcome, we derived (i) a binary case indicator (initiation by FU4), (ii) the timing of first initiation (used to compute time from baseline to first use, in months), and (iii) a censoring time at FU4 for participants without initiation. Controls were defined as participants with no initiation through FU4 and with documented observation at the 3-year follow-up (FU3; \texttt{3\_year\_follow\_up\_y\_arm\_1}); participants without FU3 observation were coded as missing for case--control phenotypes.

\subsection{Time-varying Cox survival models}

We analyzed four outcomes (alcohol initiation, cannabis initiation, nicotine initiation, and any substance initiation) using time-varying Cox proportional hazards models~\cite{8}. The analysis used three main steps implemented: (1) build a harmonized longitudinal covariate table across study visits, (2) convert each participant’s longitudinal records into start–stop (counting-process) intervals up to their event/censoring time, and (3) fit univariate and multivariable Cox models with robust standard errors.

\subsubsection{Time scale, outcomes, and overall goal}

We studied time until first substance-use initiation (four outcomes: alcohol, cannabis, nicotine, and any substance). For each outcome, every participant had: a follow-up time (the month when they first reported the outcome, or the last month they were observed), and an event indicator (1 = the event happened by that time, 0 = they were censored, meaning we did not observe the event before follow-up ended).

Time was measured in months since the participant’s baseline visit.

Our goal was to link time-varying risk factors (things that can change across visits, like environment or behavior measures) and time-invariant factors (like polygenic risk scores, PRS) to the hazard of initiation (the instantaneous risk of the event at a given time, among those who have not yet initiated).

\subsubsection{Building the time-varying covariate dataset}

Participants were assessed at multiple visits (baseline and follow-ups), with many predictors stored across separate tables. We constructed a unified longitudinal dataset through the following steps:

\begin{itemize}
    \item \textbf{Stacking visits:} For each predictor, we stacked observations across all available visits (baseline and follow-ups) into a long-format structure.
    
    \item \textbf{Merging predictors:} Predictors from different sources were merged using participant identifiers and visit labels, such that each row corresponded to a single participant at a single visit.
    
    \item \textbf{Time alignment:} We defined a common time axis using interview age. Interview age was converted to months when necessary, and follow-up time was computed as
    \[
    \texttt{time\_months} = \texttt{interview\_age\_months} - \texttt{baseline\_age\_months},
    \]
    so that baseline corresponded to time 0 for all participants.
    
    \item \textbf{Adjustment variables:} Standard covariates (sex, age, study site, and genetic principal components) were added to all records.
    
    \item \textbf{Predictor quality control:} We performed basic filtering of predictors by removing (i) variables that were entirely missing, (ii) variables with near-zero variance, and (iii) variables that were highly collinear (i.e., near-duplicate predictors with very high correlation). Sparse predictors (e.g., variables measured only at baseline) were intentionally retained.
    
    \item \textbf{Integration of PRS:} Polygenic risk scores (PRS) were merged by participant identifier. Although PRS are time-invariant, they were repeated across all visit-level rows for each participant so they could be incorporated alongside other predictors.
\end{itemize}

\subsubsection{Construction of Start--Stop (Counting Process) Intervals}

Cox models with time-varying predictors require data in interval (``counting process'') form, where each participant contributes multiple rows. Each row represents a time interval $[\text{start}, \text{stop})$, predictors are assumed constant within the interval (taking values at the beginning of the interval), and events can occur only at the end of an interval.

For each participant and each outcome, intervals were constructed as follows:

\begin{itemize}
    \item \textbf{Restrict to observed follow-up:} We retained only visits with $\texttt{time\_months} \leq$ the participant-specific follow-up time for the given outcome.
    
    \item \textbf{Order visits by time:} Visits were sorted in ascending order of $\texttt{time\_months}$.
    
    \item \textbf{Ensure a common baseline (time 0):} If the first observed visit occurred after month 0, we created a baseline row at time 0 by copying the earliest available observation, ensuring all participants share a common time origin.
    
    \item \textbf{Create consecutive intervals:} Let the ordered visit times be $t_1, t_2, \dots, t_n$. Intervals were defined as:
    \[
    \text{start}_i =
    \begin{cases}
    0, & i = 1, \\
    t_i, & i > 1,
    \end{cases}
    \qquad
    \text{stop}_i =
    \begin{cases}
    t_{i+1}, & i < n, \\
    T, & i = n,
    \end{cases}
    \]
    where $T$ denotes the participant-specific follow-up time. The interval-level event indicator was defined as $\text{event\_interval} = 1$ only for the final interval if the participant experienced the event, and $0$ otherwise.
    
    \item \textbf{Remove invalid intervals:} Intervals with $\text{stop} \leq \text{start}$ were excluded.
\end{itemize}

This procedure yields a standard Cox start--stop dataset, in which each participant contributes one or more intervals up to the time of event or censoring.

\subsubsection{Univariate Time-Varying Cox Models}

For each outcome (alcohol, nicotine, cannabis, and any substance initiation), we fitted separate univariate Cox proportional hazards models for each candidate predictor while adjusting for standard covariates.

Let $X_i(t)$ denote the value of a given predictor for participant $i$ at time $t$, which may vary over time. Let $\mathbf{C}_i$ denote the vector of adjustment covariates, including sex, age at baseline, recruitment site, and the first 20 genetic principal components.

The hazard function for participant $i$ at time $t$ was specified as:
\[
h_i(t \mid X_i(t), \mathbf{C}_i) = h_0(t) \exp\left( \beta X_i(t) + \boldsymbol{\gamma}^\top \mathbf{C}_i \right),
\]
where $h_0(t)$ is the baseline hazard function, $\beta$ is the log hazard ratio associated with the predictor $X_i(t)$, and $\boldsymbol{\gamma}$ is the vector of coefficients for the adjustment covariates. The hazard ratio (HR) for the predictor is given by $\exp(\beta)$.

Models were fitted using the start--stop (counting process) formulation implemented via \texttt{Surv(start, stop, event\_interval)}. Robust standard errors clustered at the participant level were used to account for within-person correlation arising from repeated intervals.

From each model, we reported the effect estimate and associated uncertainty, including the hazard ratio (HR), 95\% confidence interval, and corresponding $p$-value, along with the number of participants and observed events.

\subsubsection{Multivariable Time-Varying Cox Models with Data-Driven Selection}

For each outcome, we fitted a multivariable time-varying Cox proportional hazards model after data-driven predictor selection. The final model used the same start--stop (counting process) interval structure and included (i) forced adjustment covariates and (ii) a subset of selected predictors. Study site was incorporated via stratification, allowing the baseline hazard to vary across sites.

Let $\mathbf{X}_i(t)$ denote the vector of selected predictors for participant $i$ at time $t$, and let $\mathbf{Z}_i$ denote the vector of forced adjustment covariates. The hazard function was specified as:
\[
h_i(t \mid \mathbf{X}_i(t), \mathbf{Z}_i, \text{site}_i) 
= h_{0,\text{site}_i}(t)\, \exp\left( \boldsymbol{\beta}^\top \mathbf{X}_i(t) + \boldsymbol{\gamma}^\top \mathbf{Z}_i \right),
\]
where $h_{0,\text{site}_i}(t)$ is the site-specific baseline hazard, $\boldsymbol{\beta}$ are coefficients for selected predictors, and $\boldsymbol{\gamma}$ are coefficients for adjustment covariates. The forced covariates $\mathbf{Z}_i$ included sex, age at baseline, and genetic principal components (PC1--PC20). Robust standard errors clustered at the participant level were used to account for within-person correlation.

\paragraph{Preventing time leakage}
Some predictors were not measured at every visit and may only appear later during follow-up. To avoid incorporating future information, we applied a forward-fill (last observation carried forward) strategy within each participant, such that observed values were propagated to subsequent intervals. No back-filling into earlier time points was performed.

\paragraph{Handling missingness}
For numeric predictors with intermittent missingness, we adopted a ``value + availability'' representation. Specifically, for each predictor $X$, we included (i) an indicator variable $X_{\text{obs}}$ (1 if observed, 0 otherwise), and (ii) an imputed value where missing entries were replaced with the median of observed values. This approach allows the model to distinguish between low values and unavailable measurements, while retaining predictors that begin to be collected later in follow-up.

\paragraph{Pre-selection of predictors}
To reduce dimensionality and avoid extremely sparse predictors, we applied two filters prior to model selection: (i) a univariate screening step retaining predictors showing evidence of association with the outcome, and (ii) a coverage filter requiring that a predictor be observed at least once in at least 50\% of participants. Forced covariates (sex, age, site, and PCs) were always retained for confounding control and were not subject to selection.

\paragraph{LASSO-based selection}
We applied cross-validated LASSO to select a subset of predictors for each outcome. Because standard LASSO implementations do not fully accommodate start--stop risk sets, LASSO was used as a screening step based on interval end times. Predictors with non-zero coefficients at the selected penalty level were retained for the final model.

\paragraph{Final multivariable Cox model}
Using the selected predictors, we fitted the final counting-process Cox model with \texttt{Surv(start, stop, event\_interval)} as the outcome. The model included selected predictors (and corresponding availability indicators where applicable) and forced adjustment covariates (sex, age, site, and PCs). Robust standard errors clustered by participant were used. We reported hazard ratios (HRs), 95\% confidence intervals, $p$-values, and the number of participants and events.

\subsection{Causal Inference Analysis: Marginal Structural Models with IPTW}

To estimate the causal effects of time-varying exposures on the risk of substance use initiation, we applied marginal structural models (MSMs) with inverse probability of treatment weighting (IPTW) using the start--stop interval dataset. Each participant contributed multiple rows, corresponding to follow-up intervals, with an indicator for whether initiation occurred at the end of each interval. Analyses were conducted separately for alcohol, nicotine, cannabis, and any substance initiation. Baseline covariates included sex, age at baseline, recruitment site, and genetic principal components.

\paragraph{Stabilized weights}
Let $A_{ik}$ denote the exposure status for participant $i$ at interval $k$, $Y_{ik}$ the event indicator at the end of the interval, $\mathbf{Z}_i$ baseline covariates, and $\mathbf{L}_{ik}$ time-varying confounders. Stabilized inverse probability weights were defined as:
\[
SW_{ik} = \prod_{m=1}^{k} 
\frac{P\!\left(A_{im} \mid A_{i,m-1}, \mathbf{Z}_i, t_{im}\right)}
     {P\!\left(A_{im} \mid A_{i,m-1}, \mathbf{Z}_i, t_{im}, \mathbf{L}_{im}\right)}.
\]

\paragraph{Weighted outcome model}
The causal effect was estimated using a weighted pooled logistic regression model:
\[
\operatorname{logit}\!\left\{ P(Y_{ik} = 1) \right\}
= \alpha(t_{ik}) + \psi A_{ik} + \boldsymbol{\eta}^\top \mathbf{Z}_i,
\]
fitted using weights $SW_{ik}$, where $\alpha(t)$ is a flexible function of time and $\psi$ is the causal log-odds ratio associated with exposure.

\paragraph{Exposure definition and missingness handling}
For each candidate exposure, we defined a binary treatment variable at each interval. Binary variables were used directly, whereas continuous variables were dichotomized using a median split among observed values. To handle missing or late-measured variables without introducing future information, we adopted a ``value + availability'' representation: an indicator variable (1 = observed, 0 = missing) was included, and missing values were imputed using the median of observed values.

\paragraph{Estimation of IPTW}
Stabilized weights were estimated using logistic regression models for treatment assignment. The numerator model included prior treatment history, baseline covariates, and flexible functions of time. The denominator model additionally included a selected set of time-varying covariates as potential confounders.

\paragraph{Weight diagnostics and stabilization}
To mitigate bias due to violations of the positivity assumption and extreme weights, we excluded exposures with very low or very high prevalence, clipped estimated treatment probabilities away from 0 and 1, truncated weights at an upper quantile, and applied predefined weight-quality criteria prior to outcome modeling.

\paragraph{Inference}
Exposure effects were estimated using weighted pooled logistic regression over intervals, modeling the probability of initiation at each interval end. Participant-level clustered robust standard errors were used to account for repeated measures. For each outcome, only results passing weight-stability and numerical diagnostics were retained. Multiple testing across exposures was controlled using the Benjamini--Hochberg false discovery rate (FDR), and we report odds ratios (ORs), 95\% confidence intervals, $p$-values, and FDR-adjusted $q$-values.

\section{Results and Discussion}

\subsection{Study Cohort Characteristics}

We used data from the Adolescent Brain Cognitive Development (ABCD) Study\textsuperscript{\textregistered} release 5.1 \cite{5}. Participants were included if they had genome-wide genotyping data passing quality control, at least one assessment of substance use, and complete data for core covariates (sex, age, study site, and ancestry principal components). The original full analytic sample comprised 11,868 children from the ABCD baseline cohort (mean baseline age $9.91 \pm 0.62$ years). Longitudinal follow-up data were available for approximately four years. Substance-use initiation was defined separately for alcohol, nicotine, cannabis, and any substance \cite{11}. Time-to-event variables were constructed as months from baseline to first reported use, with participants censored at their last available follow-up if no initiation was reported.

For the analyses, we restricted the analytic sample to unrelated participants of European genetic ancestry with available polygenic risk scores, covariates, and longitudinal substance-use initiation information. This restriction was applied to reduce potential confounding from population stratification and genetic relatedness, and to ensure consistency between the target sample and the European-ancestry PRS construction framework.

The final EUR-unrelated analytic cohort included 2,366 participants. At baseline, participants were approximately 9.49 years old on average, with a standard deviation of 0.51 years. Assuming the ABCD sex coding of 1 = male and 2 = female, the cohort included 1,247 male participants and 1,119 female participants, corresponding to 52.7\% male and 47.3\% female.

Four time-to-event outcomes were analyzed: alcohol initiation, nicotine initiation, cannabis initiation, and any substance initiation. During follow-up, 964 participants initiated alcohol use, 151 initiated nicotine use, 100 initiated cannabis use, and 1,027 initiated any substance use. The corresponding event rates were 40.7\% for alcohol, 6.4\% for nicotine, 4.2\% for cannabis, and 43.4\% for any substance initiation.

\begin{figure*}[htbp]
\centering
\includegraphics[width=\textwidth]{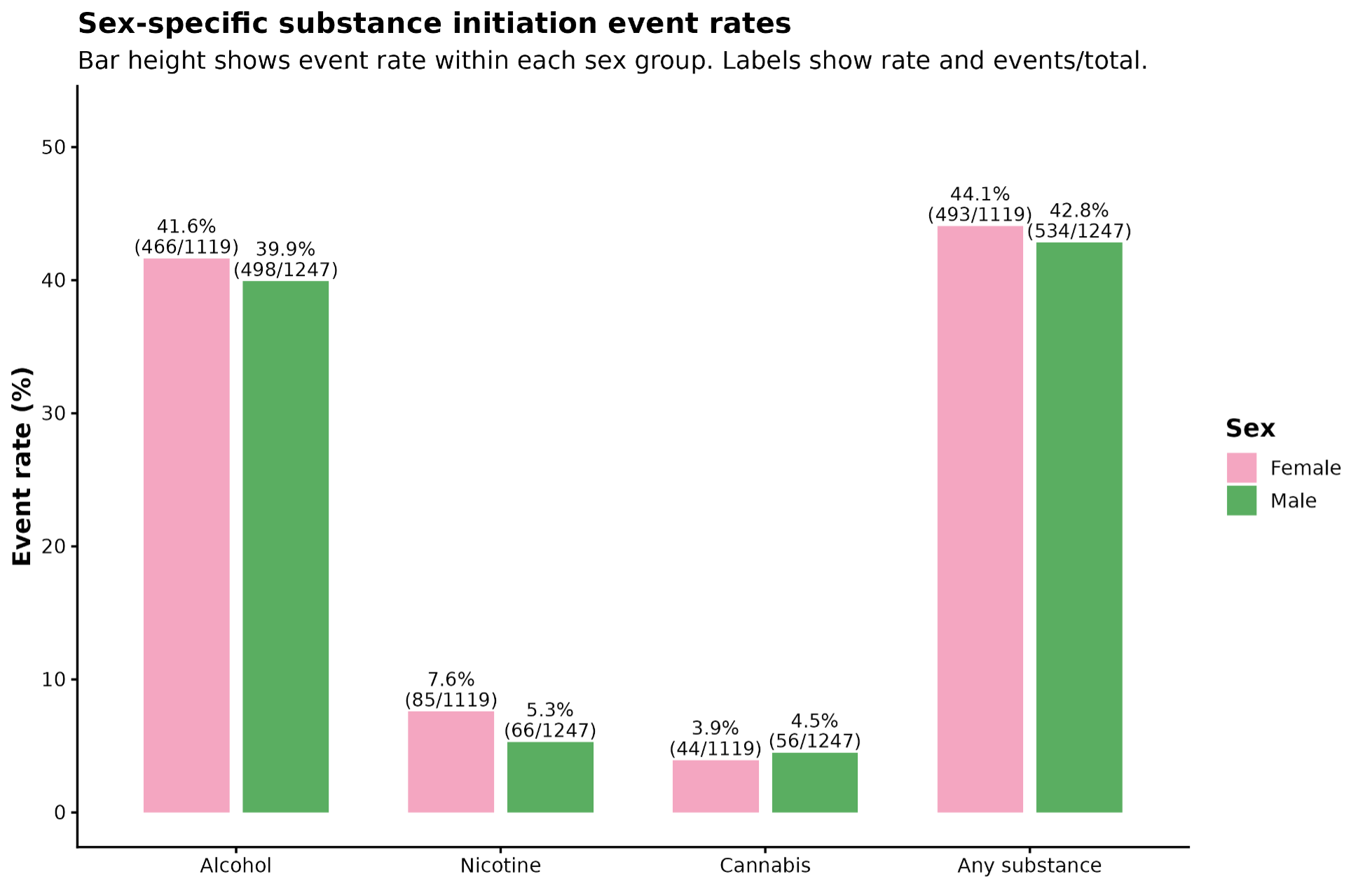}
\caption{Sex-specific substance initiation event rates among EUR-unrelated participants. Event rates were calculated separately within female and male participants in the final EUR-unrelated analytic cohort. Labels indicate the event percentage and the number of participants with observed initiation events over the sex-specific denominator. Alcohol and any substance initiation were the most common outcomes, whereas nicotine and cannabis initiation were less frequent during follow-up.}
\label{Figure2}
\end{figure*}

Sex-specific event rates were broadly similar for alcohol, cannabis, and any substance initiation, while nicotine initiation appeared somewhat higher among female participants (Figure ~\ref{Figure2}). Specifically, alcohol initiation occurred in 41.6\% of female participants and 39.9\% of male participants; nicotine initiation occurred in 7.6\% of female participants and 5.3\% of male participants; cannabis initiation occurred in 3.9\% of female participants and 4.5\% of male participants; and any substance initiation occurred in 44.1\% of female participants and 42.8\% of male participants.

Detailed distributions of initiation timing, as well as stratified summaries by sex and race/ethnicity, are provided in the Supplementary Materials.

\begin{table*}[htbp]
\centering
\caption{Sex-specific substance initiation event rates in the EUR-unrelated analytic cohort. Event rates are shown within each sex group. The denominator was 1,119 for female participants and 1,247 for male participants. Overall rates were calculated using the full EUR-unrelated analytic cohort of 2,366 participants.}
\small
\setlength{\tabcolsep}{3pt}
\begin{tabularx}{\textwidth}{
    >{\RaggedRight\arraybackslash}p{2.7cm}
    >{\centering\arraybackslash}X
    >{\centering\arraybackslash}X
    >{\centering\arraybackslash}X
    >{\centering\arraybackslash}X
    >{\centering\arraybackslash}X
    >{\centering\arraybackslash}X}
\bottomrule
Outcome &
Overall events / $N$ &
Overall event rate &
Female events / $N$ &
Female event rate &
Male events / $N$ &
Male event rate \\
\bottomrule
Alcohol initiation       & 964 / 2,366   & 40.7\% & 466 / 1,119 & 41.6\% & 498 / 1,247 & 39.9\% \\
Nicotine initiation      & 151 / 2,366   & 6.4\%  & 85 / 1,119  & 7.6\%  & 66 / 1,247  & 5.3\%  \\
Cannabis initiation      & 100 / 2,366   & 4.2\%  & 44 / 1,119  & 3.9\%  & 56 / 1,247  & 4.5\%  \\
Any substance initiation & 1,027 / 2,366 & 43.4\% & 493 / 1,119 & 44.1\% & 534 / 1,247 & 42.8\% \\
\bottomrule
\end{tabularx}
\end{table*}

\subsection{Univariate Time-Varying Survival Analysis Results}

We first used univariate time-varying Cox survival models to screen predictors associated with substance use initiation. Four outcomes were examined separately: alcohol, nicotine, cannabis, and any substance use initiation. Each predictor was tested in a separate model. All models were adjusted for age, sex, study site, and genetic principal components. Multiple testing was controlled using false discovery rate correction (FDR) within each outcome.

\begin{figure*}[htbp]
\centering
\includegraphics[width=\textwidth]{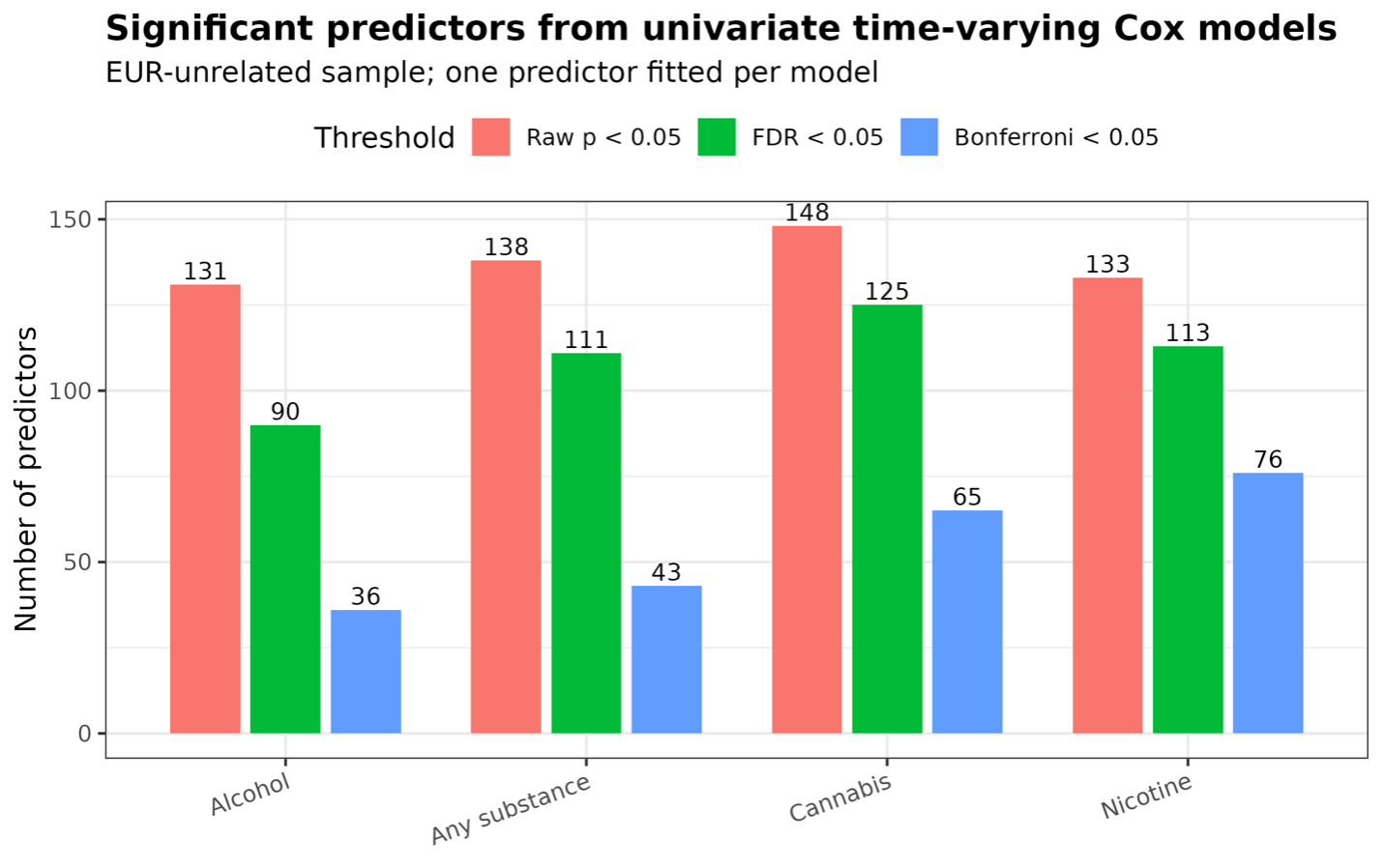}
\caption{Number of FDR-significant predictors from univariate time-varying Cox survival models. This figure shows the number of predictors significantly associated with each substance use initiation outcome after FDR correction. Each predictor was tested separately in a covariate-adjusted time-varying Cox model. The results summarize the overall burden of univariate associations for alcohol, nicotine, cannabis, and any substance use initiation. A larger number of significant predictors indicates that initiation for that outcome was associated with a broader set of time-varying factors.}
\label{Figure3}
\end{figure*}

Across the four outcomes, many predictors were associated with substance use initiation after FDR correction (Figure~\ref{Figure3}). Among 349--351 tested predictors per outcome, the number of FDR-significant predictors was 90 for alcohol, 113 for nicotine, 125 for cannabis, and 111 for any substance use initiation. A smaller number of predictors remained significant after Bonferroni correction, including 36 for alcohol, 76 for nicotine, 65 for cannabis, and 43 for any substance use initiation. These results suggest that adolescent substance use initiation is associated with a broad set of time-varying factors rather than a single predictor.

\begin{figure*}[htbp]
\centering
\includegraphics[width=\textwidth]{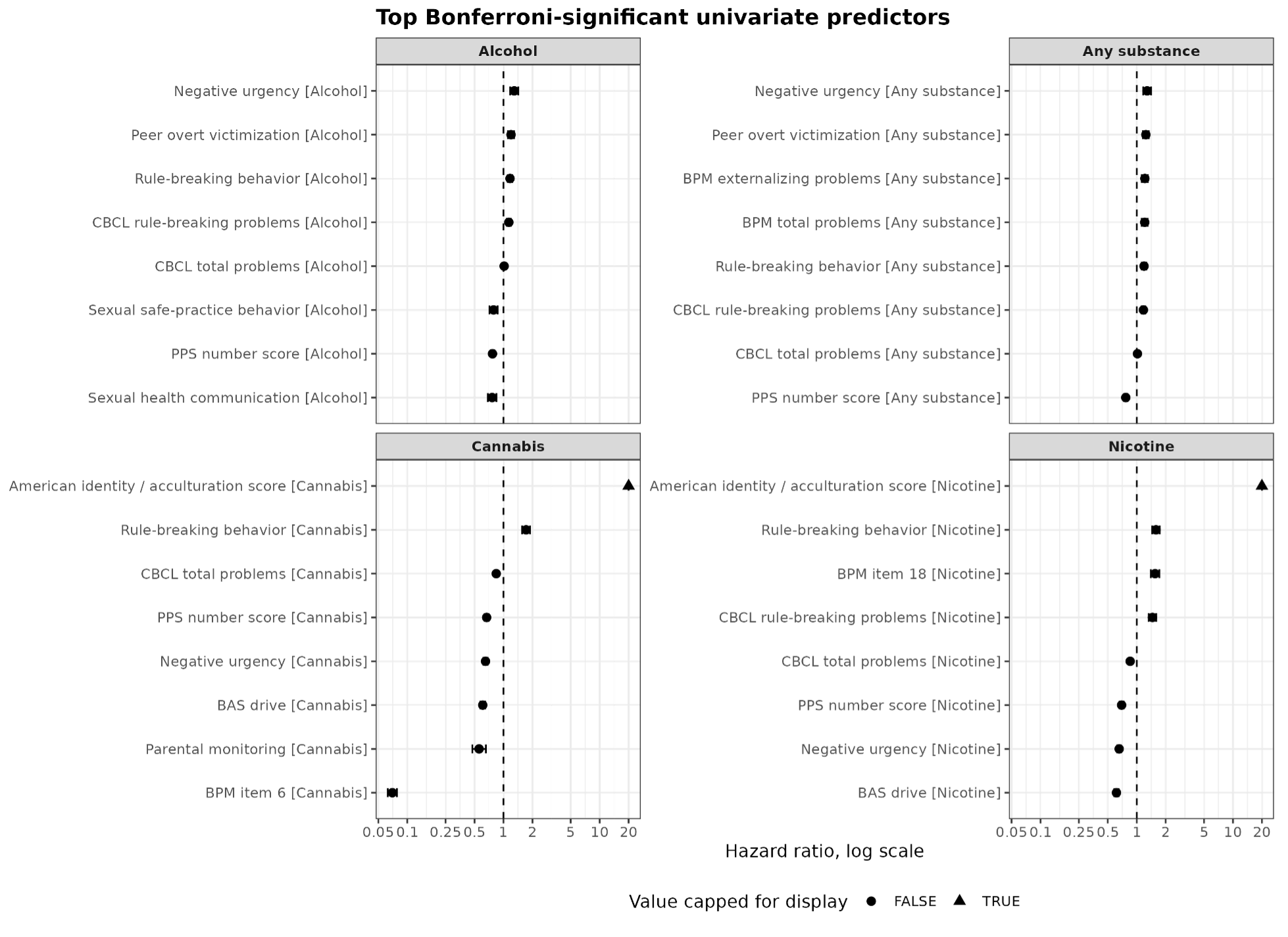}
\caption{Top Bonferroni-significant predictors from univariate time-varying Cox survival models. The figure shows the top eight Bonferroni-significant predictors for each substance use initiation outcome, ranked by FDR-adjusted significance. Each point represents the hazard ratio from a separate covariate-adjusted time-varying Cox model. Hazard ratios greater than 1 indicate higher initiation risk or earlier initiation, while hazard ratios less than 1 indicate lower initiation risk or delayed initiation. The x-axis is shown on a log scale. Extreme values were capped only for visualization and are marked in the legend. Variable names were shortened to improve readability.}
\label{Figure4}
\end{figure*}

The significant predictors covered multiple domains, including behavioral, psychosocial, family, environmental, and lifestyle-related factors (Figure~\ref{Figure4}). In general, predictors associated with higher initiation risk included poorer family environment, more behavioral or emotional problems, greater exposure to adverse social contexts, and weaker supervision or support. Predictors associated with lower initiation risk generally reflected more supportive family relationships, stronger monitoring, better school or social functioning, and lower levels of externalizing or risk-related behaviors. Full results are provided in the Supplementary Materials.

Hazard ratios were used to describe the direction of association. A hazard ratio greater than 1 indicates that higher values of the predictor were associated with higher risk or earlier initiation. A hazard ratio less than 1 indicates that higher values of the predictor were associated with lower risk or delayed initiation.

Polygenic risk score associations (PRS) were weaker and less consistent than the non-genetic predictors. In the FDR-significant results, \texttt{PRS\_CanUD} was associated with any substance use initiation and cannabis initiation, and \texttt{PRS\_nicotine} was associated with cannabis initiation. However, these PRS associations did not remain significant after Bonferroni correction. Therefore, the PRS results should be interpreted cautiously in this EUR-unrelated time-varying survival analysis.

Overall, the univariate survival analysis shows that substance use initiation is linked to multiple time-varying developmental factors. These findings support a multi-domain view of initiation risk involving family, environment, behavior, and psychosocial context. However, because each predictor was tested separately, these results should be interpreted as screening-level associations. They do not show independent or causal effects. The significant predictors were therefore used to summarize broad risk domains and to guide later multivariable and causal modeling analyses.

\subsection{Multivariable Time-Varying Survival Models}

We next performed LASSO-selected multivariable time-varying Cox proportional hazards models to identify predictors associated with substance use initiation after adjustment for demographic and genetic covariates. All models included age, sex, and the first 20 PCs as covariates, with study site included as a stratification factor. Robust standard errors were estimated using participant-level clustering. Continuous predictors were standardized, so hazard ratios represent the relative change in initiation hazard per 1-SD increase in the predictor.

Across all models, the analytic sample included 2,366 participants and 19,446 time-varying intervals. The number of initiation events differed by outcome, with 964 alcohol initiation events, 100 cannabis initiation events, 151 nicotine initiation events, and 1,027 any substance use initiation events. After LASSO selection, the final multivariable models retained 51 predictors for alcohol initiation, 30 predictors for cannabis initiation, 36 predictors for nicotine initiation, and 43 predictors for any substance use initiation.

After false discovery rate correction within each outcome, 14 predictors remained significant for alcohol initiation, 1 predictor remained significant for cannabis initiation, 15 predictors remained significant for nicotine initiation, and 3 predictors remained significant for any substance use initiation. Bonferroni correction was more conservative, retaining 4 predictors for alcohol initiation, 1 predictor for cannabis initiation, 7 predictors for nicotine initiation, and 2 predictors for any substance use initiation.

For alcohol initiation, several predictors remained significant after FDR correction. Higher financial adversity was associated with increased alcohol initiation hazard, whereas larger household size was associated with lower hazard. Parent-reported alcohol-related problems were also associated with increased alcohol initiation hazard. Youth rule-breaking behavior, parent perseverative problems, teacher-reported total problems, lack of planning, sensation seeking, family expression, family income category, and parent problematic media use were positively associated with alcohol initiation. In contrast, higher parental monitoring or cultural value score, better adaptive functioning with friends, and teacher-reported internalizing problems were associated with lower alcohol initiation hazard.

For cannabis initiation, youth rule-breaking behavior was the only predictor that remained significant after FDR and Bonferroni correction. Higher rule-breaking behavior was associated with increased cannabis initiation hazard.

For nicotine initiation, the final model identified the largest number of FDR-significant predictors. Youth rule-breaking behavior, youth phone use, lifetime events affective burden, romantic relationship experience, and several resiliency-related measures were associated with increased nicotine initiation hazard. In contrast, parent-reported total problems, parent intrusive behavior, peer reputation/aggression-related measures, weekend screen time reported by parents, peer monitoring, and selected youth behavior problem measures were associated with lower nicotine initiation hazard. Seven nicotine predictors also survived Bonferroni correction, supporting nicotine initiation as the outcome with the most robust multivariable signal in this analysis.

For any substance use initiation, three predictors remained significant after FDR correction. Parent-reported alcohol-related problems, financial adversity, and youth rule-breaking behavior were associated with increased hazard of initiating any substance use. Parent-reported alcohol-related problems and financial adversity also survived Bonferroni correction. These findings suggest that a smaller set of shared predictors may contribute to general substance use initiation, while alcohol-, cannabis-, and nicotine-specific models showed more outcome-specific patterns.

Overall, the multivariable survival results suggest that youth behavioral dysregulation, family socioeconomic adversity, parent substance-related problems, screen- or media-related factors, and selected family- or peer-context measures are associated with substance use initiation during follow-up.

\begin{table*}[!t]
\centering
\scriptsize
\setlength{\tabcolsep}{2pt}
\renewcommand{\arraystretch}{1.1}

\caption{FDR-significant predictors from LASSO-selected multivariable
time-varying Cox models for substance use initiation.}
\label{tab:multivariable-cox-results}

\resizebox{\textwidth}{!}{%
\begin{tabular}{
    >{\RaggedRight\arraybackslash}p{2.0cm}
    >{\RaggedRight\arraybackslash}p{3.8cm}
    >{\RaggedRight\arraybackslash}p{3.1cm}
    >{\RaggedRight\arraybackslash}p{1.6cm}
    >{\centering\arraybackslash}p{1.0cm}
    >{\centering\arraybackslash}p{2.4cm}
    >{\centering\arraybackslash}p{1.1cm}
    >{\centering\arraybackslash}p{1.1cm}
    >{\centering\arraybackslash}p{1.4cm}
    >{\centering\arraybackslash}p{1.0cm}
    >{\centering\arraybackslash}p{1.3cm}
}
\bottomrule
Outcome
& Predictor
& Domain
& Direction
& $\beta$
& HR (95\% CI)
& Raw $p$
& FDR $p$
& Bonferroni $p$
& Events
& Participants \\
%\midrule
\bottomrule

Alcohol initiation
& Parent-reported MACV scale score
& Cultural/family context
& Lower hazard
& $-0.162$
& 0.851 (0.783, 0.924)
& $<0.001$
& 0.006
& 0.006
& 964
& 2,366 \\

Alcohol initiation
& Financial adversity
& Socioeconomic context
& Higher hazard
& 0.126
& 1.134 (1.054, 1.219)
& $<0.001$
& 0.010
& 0.036
& 964
& 2,366 \\

Alcohol initiation
& Parent alcohol-related problems
& Parent substance-related problems
& Higher hazard
& 0.080
& 1.084 (1.034, 1.136)
& $<0.001$
& 0.010
& 0.040
& 964
& 2,366 \\

Alcohol initiation
& Household size
& Household context
& Lower hazard
& $-0.128$
& 0.880 (0.816, 0.949)
& $<0.001$
& 0.010
& 0.046
& 964
& 2,366 \\

Alcohol initiation
& Parent perseverative problems
& Parent mental health
& Higher hazard
& 0.092
& 1.097 (1.038, 1.159)
& $<0.001$
& 0.010
& 0.050
& 964
& 2,366 \\

Alcohol initiation
& Youth rule-breaking behavior
& Youth behavior
& Higher hazard
& 0.064
& 1.066 (1.021, 1.113)
& 0.004
& 0.031
& 0.187
& 964
& 2,366 \\

Alcohol initiation
& Teacher-reported total problems
& Teacher-reported behavior
& Higher hazard
& 0.126
& 1.134 (1.036, 1.242)
& 0.007
& 0.044
& 0.332
& 964
& 2,366 \\

Alcohol initiation
& Parent-reported adaptive functioning with friends
& Social/adaptive functioning
& Lower hazard
& $-0.076$
& 0.927 (0.876, 0.981)
& 0.009
& 0.044
& 0.435
& 964
& 2,366 \\

Alcohol initiation
& Teacher-reported internalizing problems
& Teacher-reported behavior
& Lower hazard
& $-0.113$
& 0.893 (0.819, 0.974)
& 0.010
& 0.044
& 0.520
& 964
& 2,366 \\

Alcohol initiation
& Youth lack of planning
& Youth impulsivity
& Higher hazard
& 0.102
& 1.107 (1.024, 1.197)
& 0.010
& 0.044
& 0.521
& 964
& 2,366 \\

Alcohol initiation
& Family expression score
& Family environment
& Higher hazard
& 0.063
& 1.065 (1.015, 1.118)
& 0.011
& 0.044
& 0.558
& 964
& 2,366 \\

Alcohol initiation
& Family income category
& Socioeconomic context
& Higher hazard
& 0.097
& 1.101 (1.022, 1.186)
& 0.011
& 0.044
& 0.561
& 964
& 2,366 \\

Alcohol initiation
& Youth sensation seeking
& Youth impulsivity
& Higher hazard
& 0.097
& 1.102 (1.022, 1.187)
& 0.011
& 0.044
& 0.571
& 964
& 2,366 \\

Alcohol initiation
& Parent problematic media use score
& Technology/media environment
& Higher hazard
& 0.069
& 1.072 (1.015, 1.131)
& 0.012
& 0.044
& 0.610
& 964
& 2,366 \\

\bottomrule

Any substance use initiation
& Parent alcohol-related problems
& Parent substance-related problems
& Higher hazard
& 0.088
& 1.092 (1.044, 1.142)
& $<0.001$
& 0.005
& 0.005
& 1,027
& 2,366 \\

Any substance use initiation
& Financial adversity
& Socioeconomic context
& Higher hazard
& 0.117
& 1.125 (1.053, 1.201)
& $<0.001$
& 0.010
& 0.020
& 1,027
& 2,366 \\

Any substance use initiation
& Youth rule-breaking behavior
& Youth behavior
& Higher hazard
& 0.070
& 1.073 (1.028, 1.119)
& 0.001
& 0.017
& 0.051
& 1,027
& 2,366 \\

\bottomrule

Cannabis initiation
& Youth rule-breaking behavior
& Youth behavior
& Higher hazard
& 0.205
& 1.228 (1.126, 1.340)
& $<0.001$
& $<0.001$
& $<0.001$
& 100
& 2,366 \\

\bottomrule

Nicotine initiation
& Parent-reported child total problems
& Parent-reported child behavior
& Lower hazard
& $-0.094$
& 0.910 (0.891, 0.930)
& $<0.001$
& $<0.001$
& $<0.001$
& 151
& 2,366 \\

Nicotine initiation
& Youth resiliency item 7a
& Youth resilience
& Higher hazard
& 0.141
& 1.152 (1.071, 1.238)
& $<0.001$
& 0.002
& 0.005
& 151
& 2,366 \\

Nicotine initiation
& Affective burden from life events
& Life events
& Higher hazard
& 0.270
& 1.310 (1.136, 1.511)
& $<0.001$
& 0.003
& 0.008
& 151
& 2,366 \\

Nicotine initiation
& Parent intrusive behavior
& Parent mental health/behavior
& Lower hazard
& $-0.346$
& 0.707 (0.586, 0.853)
& $<0.001$
& 0.003
& 0.010
& 151
& 2,366 \\

Nicotine initiation
& Youth rule-breaking behavior
& Youth behavior
& Higher hazard
& 0.129
& 1.137 (1.059, 1.221)
& $<0.001$
& 0.003
& 0.015
& 151
& 2,366 \\

Nicotine initiation
& Peer reputation/aggression problems
& Peer context
& Lower hazard
& $-0.165$
& 0.848 (0.773, 0.931)
& $<0.001$
& 0.003
& 0.018
& 151
& 2,366 \\

Nicotine initiation
& Youth resiliency item 7b
& Youth resilience
& Lower hazard
& $-0.145$
& 0.865 (0.793, 0.944)
& 0.001
& 0.006
& 0.042
& 151
& 2,366 \\

Nicotine initiation
& Youth phone use score
& Technology/media environment
& Higher hazard
& 0.190
& 1.209 (1.076, 1.359)
& 0.001
& 0.006
& 0.050
& 151
& 2,366 \\

Nicotine initiation
& Ever had a boyfriend or girlfriend
& Romantic/social development
& Higher hazard
& 0.204
& 1.227 (1.075, 1.400)
& 0.002
& 0.010
& 0.088
& 151
& 2,366 \\

Nicotine initiation
& Parent-reported weekend screen time
& Technology/media environment
& Lower hazard
& $-0.194$
& 0.824 (0.724, 0.938)
& 0.003
& 0.012
& 0.123
& 151
& 2,366 \\

Nicotine initiation
& Youth Brief Problem Monitor item 10
& Youth behavior
& Higher hazard
& 0.251
& 1.285 (1.080, 1.529)
& 0.005
& 0.015
& 0.166
& 151
& 2,366 \\

Nicotine initiation
& Youth-reported parental monitoring score
& Parent monitoring
& Lower hazard
& $-0.223$
& 0.800 (0.676, 0.948)
& 0.010
& 0.027
& 0.353
& 151
& 2,366 \\

Nicotine initiation
& Parent-reported child rule-breaking problems
& Parent-reported child behavior
& Higher hazard
& 0.145
& 1.156 (1.036, 1.291)
& 0.010
& 0.027
& 0.355
& 151
& 2,366 \\

Nicotine initiation
& Youth resiliency mean score
& Youth resilience
& Higher hazard
& 0.215
& 1.240 (1.045, 1.472)
& 0.014
& 0.035
& 0.494
& 151
& 2,366 \\

Nicotine initiation
& Youth Brief Problem Monitor item 18
& Youth behavior
& Higher hazard
& 0.159
& 1.172 (1.031, 1.331)
& 0.015
& 0.036
& 0.537
& 151
& 2,366 \\

\bottomrule
\end{tabular}%
}

\vspace{0.5em}

\begin{minipage}{\linewidth}
\footnotesize
\textit{Note.}
Hazard ratios were estimated from final multivariable time-varying Cox
proportional hazards models using start--stop intervals. Predictors were
selected using LASSO Cox regression at \texttt{lambda.min} and then refitted
in outcome-specific Cox models. Models were adjusted for age, sex, and the
first 20 genetic principal components, stratified by study site, and estimated
using robust standard errors clustered by participant. Continuous predictors
were standardized; therefore, hazard ratios represent the change in initiation
hazard per 1-SD increase in the predictor. FDR-adjusted $p$ values were
calculated within each outcome using the Benjamini--Hochberg procedure.
Only predictors significant at a within-outcome FDR-adjusted $p<0.05$ are
shown. Bonferroni-adjusted $p$ values are provided as a conservative
sensitivity measure. Readable predictor labels were manually assigned from
ABCD-style variable names.
\end{minipage}

\end{table*}

\subsection{Causal Analyses}

To further evaluate the robustness of the time-varying survival findings under a causal modeling framework, we fitted inverse probability of treatment weighted marginal structural models using pooled logistic regression. Across the four substance use initiation outcomes, the primary MSM analysis tested 14 candidate predictors for alcohol initiation, 10 for any substance use initiation, 9 for cannabis initiation, and 16 for nicotine initiation. No exposure was excluded because of positivity or weight instability. The minimum exposure prevalence ranged from 0.026 to 0.121 across outcomes, and the 99th percentile of the stabilized weights was low across all models, with maximum values ranging from 1.52 to 1.95, indicating no evidence of extreme weighting.

In the primary MSM analysis, 6 predictors were significantly associated with alcohol initiation, 4 with any substance use initiation, 5 with cannabis initiation, and 10 with nicotine initiation after FDR correction. Youth rule-breaking behavior was the most consistent predictor and was significantly associated with all four outcomes. Higher rule-breaking behavior was associated with increased odds of alcohol initiation (OR = 3.88, 95\% CI: 2.33--6.46), any substance use initiation (OR = 3.19, 95\% CI: 2.00--5.08), cannabis initiation (OR = 5.31, 95\% CI: 3.04--9.27), and nicotine initiation (OR = 7.72, 95\% CI: 4.00--14.92).

\begin{table*}[!t]
\centering
\scriptsize
\setlength{\tabcolsep}{2.5pt}
\renewcommand{\arraystretch}{1.15}

\caption{FDR-significant predictors of substance use initiation in marginal structural models.}
\label{tab:msm-results}

\resizebox{\textwidth}{!}{%
\begin{tabular}{
    >{\RaggedRight\arraybackslash}p{1.8cm}
    >{\RaggedRight\arraybackslash}p{3.0cm}
    >{\RaggedRight\arraybackslash}p{4.1cm}
    >{\RaggedRight\arraybackslash}p{1.5cm}
    >{\centering\arraybackslash}p{2.5cm}
    >{\centering\arraybackslash}p{1.2cm}
    >{\centering\arraybackslash}p{1.5cm}
    >{\centering\arraybackslash}p{1.8cm}
    >{\RaggedRight\arraybackslash}p{3.2cm}
}
\bottomrule
Outcome
& Domain
& Predictor
& Direction
& OR (95\% CI)
& $p$ value
& FDR $q$ value
& Exposure prevalence
& Sensitivity robustness \\
\bottomrule

Alcohol
& Behavioral problems
& Youth rule-breaking behavior
& Higher odds
& 3.88 (2.33--6.46)
& $<0.001$
& $<0.001$
& 0.141
& Robust in all 3 settings \\

Alcohol
& Family media environment
& Parent problematic media use
& Higher odds
& 2.02 (1.22--3.35)
& 0.006
& 0.016
& 0.253
& Robust in all 3 settings \\

Alcohol
& Impulsivity
& Lack of planning
& Higher odds
& 2.07 (1.32--3.26)
& 0.002
& 0.005
& 0.291
& Robust in all 3 settings \\

Alcohol
& Impulsivity
& Sensation seeking
& Higher odds
& 2.22 (1.45--3.41)
& $<0.001$
& 0.001
& 0.401
& Robust in all 3 settings \\

Alcohol
& Parental psychopathology
& Parental alcohol-related problems
& Higher odds
& 2.42 (1.47--3.99)
& $<0.001$
& 0.002
& 0.155
& Robust in all 3 settings \\

Alcohol
& Physical health
& Body weight
& Higher odds
& 1.90 (1.13--3.20)
& 0.015
& 0.035
& 0.497
& Robust in all 3 settings \\

\bottomrule

Any substance
& Behavioral problems
& Youth rule-breaking behavior
& Higher odds
& 3.19 (2.00--5.08)
& $<0.001$
& $<0.001$
& 0.141
& Robust in all 3 settings \\

Any substance
& Child behavioral problems
& Parent-rated child rule-breaking symptoms
& Higher odds
& 1.71 (1.12--2.60)
& 0.013
& 0.031
& 0.438
& Robust in all 3 settings \\

Any substance
& Impulsivity
& Sensation seeking
& Higher odds
& 1.80 (1.22--2.64)
& 0.003
& 0.009
& 0.401
& Robust in all 3 settings \\

Any substance
& Parental psychopathology
& Parental alcohol-related problems
& Higher odds
& 2.28 (1.43--3.62)
& $<0.001$
& 0.002
& 0.155
& Robust in all 3 settings \\

\bottomrule

Cannabis
& Behavioral problems
& Youth rule-breaking behavior
& Higher odds
& 5.31 (3.04--9.27)
& $<0.001$
& $<0.001$
& 0.141
& Robust in all 3 settings \\

Cannabis
& Impulsivity
& Sensation seeking
& Higher odds
& 2.00 (1.22--3.30)
& 0.006
& 0.016
& 0.401
& Robust in all 3 settings \\

Cannabis
& Parenting
& Parental monitoring
& Lower odds
& 0.43 (0.21--0.88)
& 0.020
& 0.043
& 0.370
& Primary only / not all sensitivity settings \\

Cannabis
& Peer/romantic experience
& Ever had a boyfriend/girlfriend
& Higher odds
& 4.77 (2.61--8.69)
& $<0.001$
& $<0.001$
& 0.125
& Robust in all 3 settings \\

Cannabis
& Physical health/caffeine
& Caffeine use in past 24 hours
& Higher odds
& 2.41 (1.45--4.01)
& $<0.001$
& 0.003
& 0.202
& Robust in all 3 settings \\

\bottomrule

Nicotine
& Behavioral problems
& Behavioral problem symptoms
& Higher odds
& 2.72 (1.55--4.78)
& $<0.001$
& 0.002
& 0.146
& Robust in all 3 settings \\

Nicotine
& Behavioral problems
& Youth rule-breaking behavior
& Higher odds
& 7.72 (4.00--14.92)
& $<0.001$
& $<0.001$
& 0.141
& Robust in all 3 settings \\

Nicotine
& Child behavioral problems
& Parent-rated child rule-breaking symptoms
& Higher odds
& 3.04 (1.83--5.03)
& $<0.001$
& $<0.001$
& 0.438
& Robust in all 3 settings \\

Nicotine
& Cultural/family values
& Family obligation values
& Lower odds
& 0.51 (0.29--0.89)
& 0.017
& 0.039
& 0.216
& Robust in all 3 settings \\

Nicotine
& Life events
& Negative life-event affect
& Higher odds
& 2.43 (1.43--4.14)
& 0.001
& 0.004
& 0.313
& Robust in all 3 settings \\

Nicotine
& Peer environment
& Peer reputation aggression
& Higher odds
& 2.34 (1.12--4.91)
& 0.024
& 0.047
& 0.042
& Primary only / not all sensitivity settings \\

Nicotine
& Peer/romantic experience
& Ever had a boyfriend/girlfriend
& Higher odds
& 4.51 (2.55--7.99)
& $<0.001$
& $<0.001$
& 0.125
& Robust in all 3 settings \\

Nicotine
& Resilience
& Resiliency
& Higher odds
& 2.33 (1.36--4.00)
& 0.002
& 0.007
& 0.044
& Robust in all 3 settings \\

Nicotine
& Screen/media use
& Youth phone screen use
& Higher odds
& 2.97 (1.81--4.87)
& $<0.001$
& $<0.001$
& 0.148
& Robust in all 3 settings \\

Nicotine
& Sleep
& Weekday sleep duration
& Higher odds
& 1.76 (1.08--2.88)
& 0.023
& 0.047
& 0.135
& Robust in all 3 settings \\

\bottomrule
\end{tabular}%
}

\vspace{0.5em}

\begin{minipage}{\textwidth}
\footnotesize
\textit{Note.}
Inverse probability of treatment weighted marginal structural models were
fitted using pooled logistic regression for each substance use initiation
outcome. The table shows predictors that remained significant after FDR
correction in the primary MSM analysis. Odds ratios indicate the association
between each time-varying predictor and subsequent initiation of alcohol, any
substance use, cannabis, or nicotine use. Exposure prevalence refers to the
observed prevalence of the predictor in the analytic sample. Sensitivity
robustness indicates whether the association remained FDR-significant across
the primary MSM analysis and both sensitivity settings.
\end{minipage}

\end{table*}

\begin{figure*}[htbp]
\centering
\includegraphics[width=\textwidth]{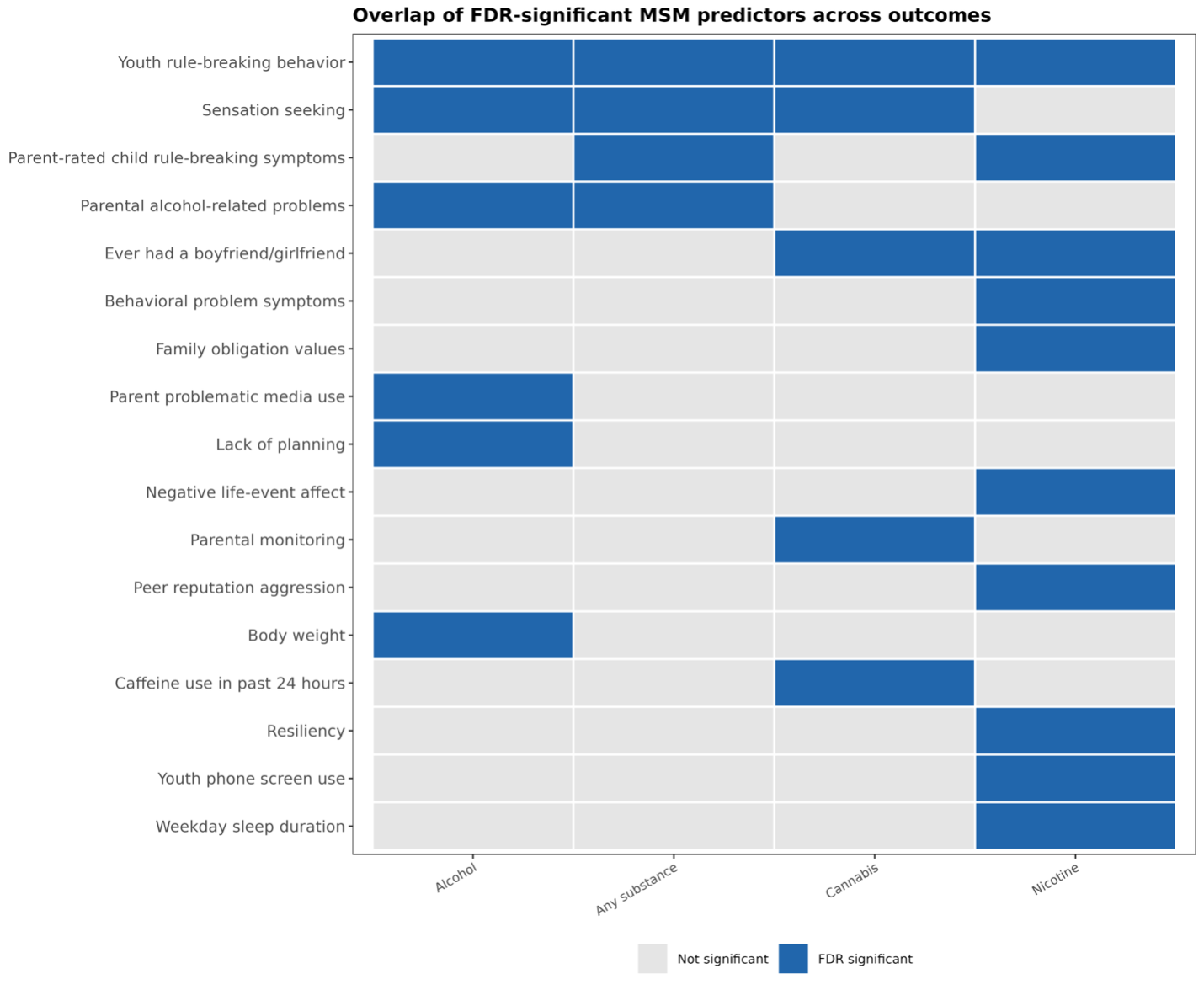}
\caption{Robust time-varying predictors of substance use initiation identified by marginal structural models. The figure summarizes FDR-significant predictor--outcome associations from the primary MSM analysis and indicates whether each association remained significant across sensitivity analyses. Youth rule-breaking behavior was the most consistent predictor, showing significant associations with alcohol, any substance use, cannabis, and nicotine initiation across all MSM settings. Other robust predictors included sensation seeking, parental alcohol-related problems, having had a boyfriend or girlfriend, and parent-rated child rule-breaking symptoms, with outcome-specific patterns across substance use initiation outcomes.}
\label{Figure5}
\end{figure*}

Several outcome-specific patterns were also observed. For alcohol initiation, significant predictors included parental alcohol-related problems, youth sensation seeking, lack of planning, parent problematic media use, body weight, and rule-breaking behavior. For any substance use initiation, significant predictors included parental alcohol-related problems, parent-rated child rule-breaking symptoms, youth sensation seeking, and youth rule-breaking behavior. For cannabis initiation, significant predictors included youth rule-breaking behavior, having had a boyfriend or girlfriend, caffeine use in the past 24 hours, sensation seeking, and lower parental monitoring. For nicotine initiation, significant predictors included youth rule-breaking behavior, having had a boyfriend or girlfriend, youth phone screen use, parent-rated child rule-breaking symptoms, behavioral problem symptoms, negative life-event affect, resiliency, cultural/family values, peer reputation aggression, and weekday sleep duration.

Sensitivity analyses supported the robustness of the primary MSM findings. Under the alternative positivity threshold setting, the number of FDR-significant predictors was identical to the primary analysis for all outcomes. Under the alternative weight truncation setting, the number of FDR-significant predictors remained similar, with 6 predictors for alcohol initiation, 4 for any substance use initiation, 4 for cannabis initiation, and 9 for nicotine initiation. In the cross-setting comparison, 23 predictor--outcome associations remained significant across the primary analysis and both sensitivity analyses. Youth rule-breaking behavior remained significant across all four substance use outcomes in every MSM setting, supporting it as the most robust time-varying predictor identified by the MSM analysis.

Because the multivariable survival models identified mainly environmental and behavioral predictors, and because PRS did not show robust independent associations after correction and model selection, we focused the subsequent marginal structural model analyses on potentially modifiable time-varying factors. This final MSM step was intended to evaluate whether the strongest observational signals from the time-varying Cox analyses remained robust under inverse probability weighting, with particular attention to predictors that may represent actionable prevention targets, such as youth rule-breaking behavior, impulsivity, parental monitoring, family risk context, media use, peer or romantic experiences, sleep, and lifestyle-related exposures.

These findings were robust to alternative propensity score clipping and truncation strategies, with stable weight diagnostics across sensitivity analyses (Figures~\ref{Figure5}).

\section{Conclusion}

In this longitudinal analysis of substance use initiation in the ABCD cohort, time-varying environmental, behavioral, family, and psychosocial factors showed substantially stronger and more robust associations with early initiation than polygenic risk scores. Across univariate and multivariable time-varying Cox models, initiation of alcohol, nicotine, cannabis, and any substance was associated with a broad set of modifiable developmental factors, including youth rule-breaking behavior, impulsivity-related traits, parental alcohol-related problems, financial adversity, family and cultural context, media or phone use, life-event burden, peer and romantic experiences, sleep-related measures, and parental monitoring.

In contrast, polygenic risk scores for alcohol, cannabis, nicotine, and general substance use disorder showed limited evidence of robust association in the EUR-unrelated analytic cohort. Although some PRS associations appeared in screening-level models, they were weaker than environmental predictors, did not consistently survive stringent multiple-testing correction, and were not prominent among the final multivariable or MSM findings. These results suggest that, in this sample and developmental window, early substance use initiation is more strongly captured by proximal time-varying environmental and behavioral factors than by measured common-variant genetic liability.

The marginal structural model analyses further supported this interpretation. Youth rule-breaking behavior emerged as the most consistent predictor, showing robust associations across alcohol, nicotine, cannabis, and any substance initiation. Additional MSM findings highlighted outcome-specific but potentially actionable factors, including sensation seeking, lack of planning, parental alcohol-related problems, parent-rated child rule-breaking symptoms, youth phone use, negative life-event affect, romantic relationship experience, caffeine exposure, sleep duration, and parental monitoring. These findings point to behavioral dysregulation, family context, peer/social development, and lifestyle-related exposures as key pathways associated with earlier substance use initiation.

Overall, the study supports a developmental model in which early adolescent substance use initiation is shaped primarily by dynamic environmental and behavioral contexts rather than by PRS alone. The findings emphasize the importance of prevention strategies targeting modifiable factors such as parental monitoring, family risk context, impulsivity, rule-breaking behavior, media use, sleep, and peer-related exposures. PRS may still be useful as background measures of genetic liability, but the current results do not support presenting PRS as a central robust predictor of initiation in this analysis.

\section{Funding}

This work was supported by the National Institutes of Health (NIH), National Institute on Drug Abuse (NIDA) under award DP1DA054373. The funder had no role in the study design; data collection, analysis, or interpretation; manuscript writing; or the decision to submit for publication. The content is solely the responsibility of the authors and does not necessarily represent the official views of the NIH.

\section{Author Contributions}
Mengman Wei conceived the study, designed the analytical framework, performed all data processing, statistical analyses, and computational modeling, and drafted the manuscript. All code implementation, data curation, and result interpretation were conducted by Mengman Wei.\\

Qian Peng provided supervision, general guidance, resource support, and funding acquisition.

\section{Preprint Notice}

This manuscript is a preprint and has not yet undergone peer review. The content is shared to disseminate findings and establish precedence. Additional analyses and revisions may be incorporated in future versions.

\section{Data availability}

\textbf{Code.} The analysis code and scripts used in this study are freely available at the following GitHub repository: \url{https://github.com/mw742/ABCD_sur_cau}.

\textbf{Data.} This study uses data from the Adolescent Brain Cognitive Development (ABCD) Study (\url{https://abcdstudy.org}), held in the NIMH Data Archive (NDA). The ABCD data release used was version 5.1. The study is supported by the National Institutes of Health (NIH) and additional federal partners under multiple award numbers, including U01DA041048 and U01DA050987. The full list of funders is available at \url{https://abcdstudy.org/federal-partners.html}.

\bibliographystyle{unsrt}   
\bibliography{reference}

\end{document}